\documentclass[10pt,pre,twocolumn,aps,showpacs,superscriptaddress]{revtex4-1}
\usepackage{epsfig,amsmath,amssymb,graphicx,color,calc,epstopdf}
\usepackage{color}
\usepackage{inputenc}

\newcommand{\br}{\mathbf{r}}
\newcommand{\bR}{\mathbf{R}}
\newcommand{\ud}{\mathrm{d}}

\newcommand{\beq}{\begin{equation}}
\newcommand{\eeq}{\end{equation}}
\newcommand{\ba}{\begin{align}}
\newcommand{\ea}{\end{align}}

\makeatletter 
\@addtoreset{equation}{section}
\makeatother  

\usepackage{xr}
\begin{document}

\title{Equilibrium Phase Behavior of a Continuous-Space Microphase Former}
\author{Yuan Zhuang}
\affiliation{Department of Chemistry, Duke University, Durham,
North Carolina 27708, USA}
\author{Kai Zhang}
\affiliation{Department of Chemical Engineering, Columbia University,
  New York, New York 10027, USA}
\author{Patrick Charbonneau}
\affiliation{Department of Chemistry, Duke University, Durham,
North Carolina 27708, USA}
\affiliation{Department of Physics, Duke University, Durham,
North Carolina 27708, USA}

\begin{abstract}
Periodic microphases universally emerge in systems for which
short-range inter-particle attraction is frustrated by long-range
repulsion. The morphological richness of these phases makes them
desirable material targets, but our relatively coarse understanding of
even simple models hinders controlling their assembly. We report here
the solution of the equilibrium phase behavior of a microscopic microphase former
through specialized Monte Carlo simulations. The results for cluster crystal, cylindrical, double
  gyroid and lamellar ordering qualitatively agree with a Landau-type free energy
description and reveal the nontrivial interplay between
cluster, gel and microphase formation.
\end{abstract}


\maketitle
\section{Introduction}
Microphases supersede simple gas-liquid coexistence when short-range
inter-particle attraction is frustrated by long-range repulsion
(SALR). The resulting structures are both elegant and remarkably
useful~\cite{Seul1995}. Block
copolymers~\cite{Bates1990,Bates1999,Kim2010}, for instance, form a
rich array of periodic structures, such as lamellae,
gyroid~\cite{Leibler1980,Matsen1996} and exotic
morphologies~\cite{Spontak1993,Epps2004,Tyler2005,Crossland2008,Scherer2013}, whose robust assembly enables industrial applications in drug
delivery~\cite{Kataoka2001,Rosler2001} and nanoscale
patterning~\cite{Li2006,Krishnamoorthy2006}, among others.
Because microphase formation constitutes a universality class of sort~\cite{Brazovskii1975},
many other systems either exhibit or share the potential to form similar
assemblies~\cite{Seul1995,Ciach2013}. In the latter category,
colloidal suspensions are particularly interesting. The relative ease
with which interactions between colloids can be tuned indeed suggests
that a broad array of ordered microphases should be
achievable~\cite{Ciach2013}. Yet, in
experiments~\cite{Campbell2005,Klix2010,Zhang2012b}  only amorphous
gels and clusters have been observed in systems ranging from proteins~\cite{Stradner2004} to micron-scale beads~\cite{Jordan2014}.

A variety of explanations have been advanced to explain the difficulty of assembling periodic microphases in colloids, including a glass-like dynamical slowdown upon approaching the microphase regime~\cite{Schmalian2000,Geissler2004}, the existence of an equilibrium gel phase~\cite{Toledano2009,Liu2005}, and the dynamical arrest of partly assembled structures due either to particle--scale sluggishness~\cite{deCandia2006,Tarzia2007,Charbonneau2007} or competition between morphologies~\cite{Zhang2006,deCandia2011,DelGado2010}. In order to obtain a clearer physical picture of these effects and thus hopefully guide
experimental microphase ordering, a better understanding of the
relationship between equilibrium statics and dynamics is
needed. Insights from theory and simulation would be 
beneficial, but both approaches face serious challenges. On the one hand, theoretical descriptions, such as
density-functional theory~\cite{Leibler1980,Bates1990,Matsen1996},
self-consistent field theory~\cite{Matsen1994}, random-phase
approximation~\cite{Imperio2004,Archer2007} and
others~\cite{Sear1999,Liu2005}, capture reasonably well the microphase
structures, but corresponding dynamical descriptions are more
limited~\cite{Geissler2004,Ruzicka2004,Tarzia2006,Tarzia2007,Berthier2011}. On
the other hand, the dynamics of particle-based
models has been extensively studied by simulations~\cite{Sciortino2004,deCandia2006,Toledano2009,Jadrich2015,Zhang2015},
but our thermodynamic grasp of these models is rather
poor~\cite{Binder2000,Mladek2007,Muller2008}. In this Letter, we
introduce the components needed to study the thermodynamic behavior of
microscopic, microphase-forming models and thus help clarify the
interplay between equilibrium ordering and sluggish dynamics.

\begin{figure*}
\includegraphics{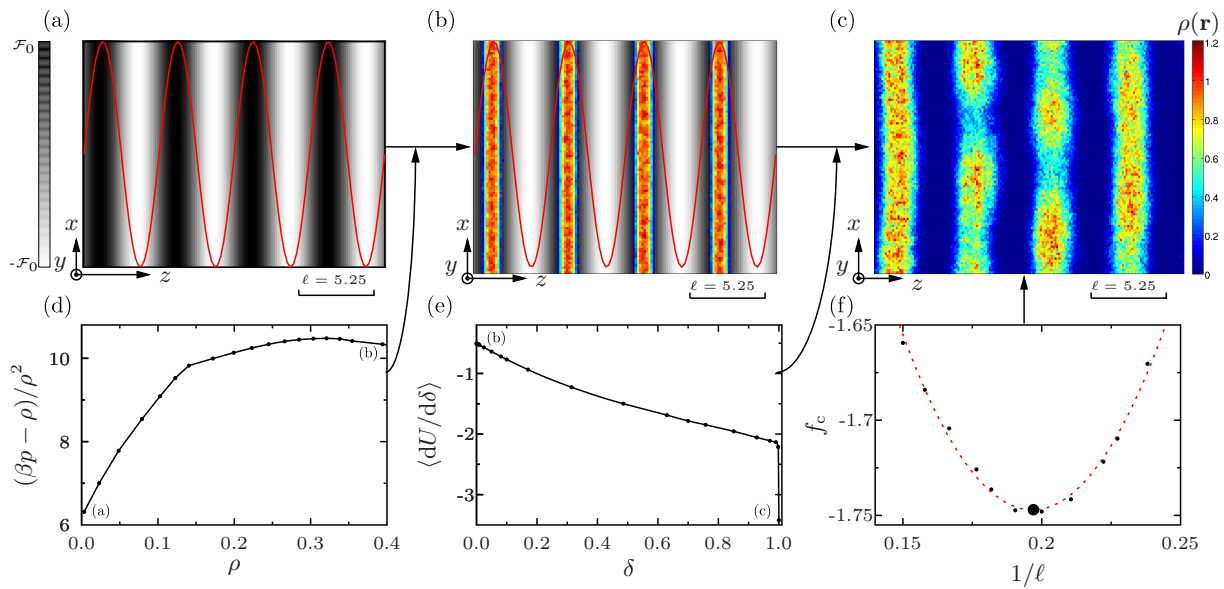}
\caption{
Two-step TI for the lamellar phase at $T=0.3$ and $\rho=0.4$.
Projections on the $xz$ plane of the coarse-grained number density $\rho(\mathbf{r})$
and external field profiles $\mathcal{F}(\br)$ for (a) $\rho=0$ with
field $\mathcal{F}(\br)=\mathcal{F}_0\cos(2\pi z/\ell)$, where
$\mathcal{F}_0=2$ and $\ell=5.25$, (b) $\rho=0.4$ with
field, and (c) $\rho=0.4$ without field. Summing the TI results
for the equation of state from (a) to (b) with those from the
alchemical transformation in $\delta$ from (b) to (c)~(see Appendix Sec.~\ref{section:TI}) -- in (d) and
(e), respectively -- gives the free energy constrained to a given area density $\varrho_{\ell}=\rho \ell$~(see Appendix Sec.~\ref{section:OccupancyAndReferenceField}).
(f) From the minimum of a quadratic fit
(dashed lines) to $f_\mathrm{c}$ (points) we obtain the
equilibrium thermodynamic $f$ at $\ell=\varrho_{\ell}/\rho=5.05$ (large dot).}\label{fig:fig1}
\end{figure*}

\begin{figure*}
 \centering
 \includegraphics[width=1\textwidth]{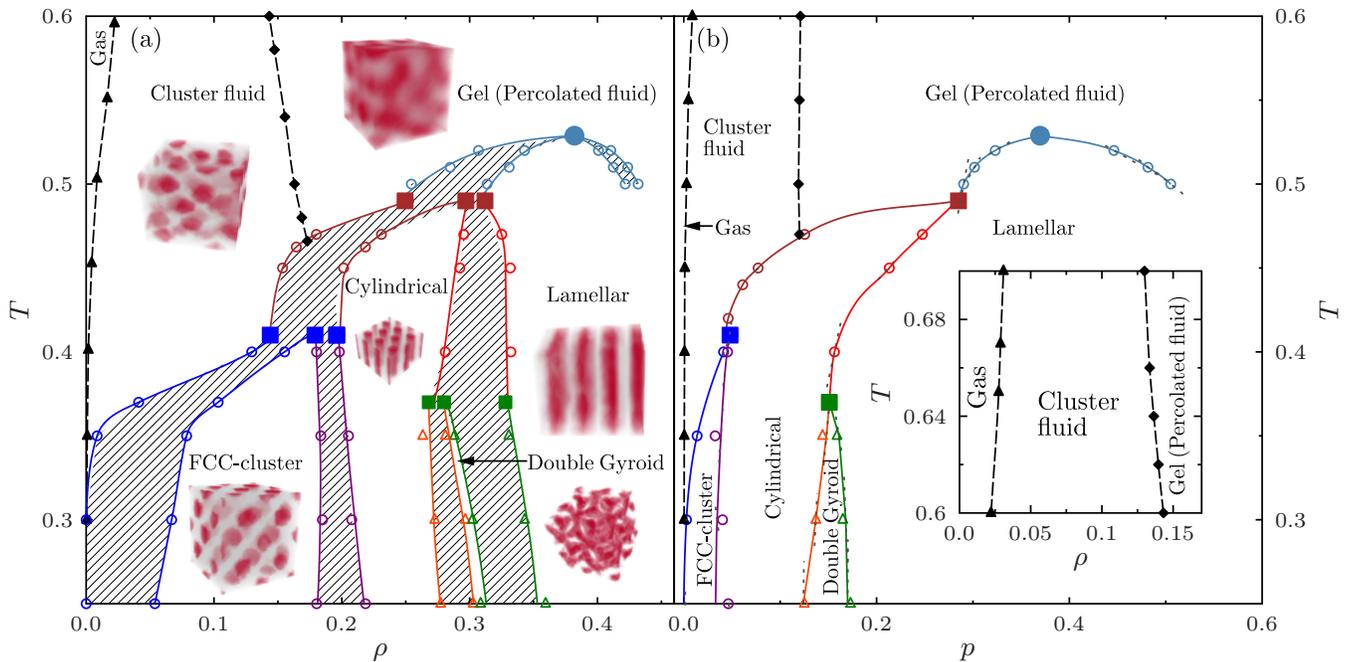}
 \caption{Summary (a) $T-\rho$ and (b) $p-T$ phase diagrams indicating the first-order (empty symbols), and the order-disorder transition (filled circle) as well as the cmc (filled triangles) and percolation (filled diamonds) lines. Errors are comparable to the symbol sizes, striped areas correspond to coexistence regime, and lines are guides for the eye.
   Clausius-Clapeyron results for the slope of the coexistence line (dashed lines) validate the numerical results in (b). Sample average density profiles for the different phases are given in (a). The inset provides percolation and cmc lines to higher $T$. Three triple points can be identified (filled squares): (i) fluid--FCC-cluster--cylindrical coexistence at $T=0.410(5)$ and $p=0.051(1)$, (ii) fluid--cylindrical--lamellar coexistence at $T=0.491(4)$
and $p=0.285(2)$, and (iii) cylindrical--double gyroid--lamellar coexistence at $T=0.37(1)$ and $p=0.15(1)$}.
 \label{fig:fig4}
 \end{figure*}

\begin{figure}[htbp]
\centering
\includegraphics[width=0.5\textwidth]{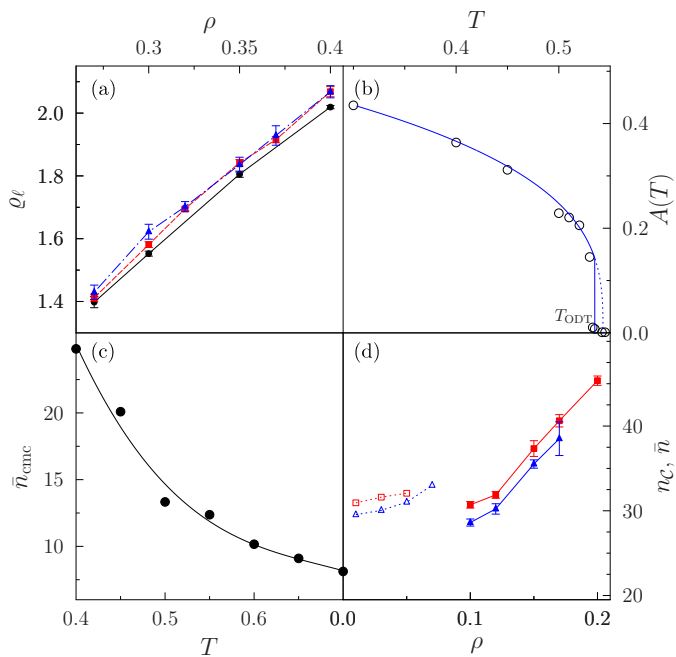}
\caption{
	(a) The equilibrium occupancy $\varrho_{\ell}$ of lamellae 
	depends only weakly on temperature in this regime, where $T=0.3$ (dots),
	$T=0.35$ (squares) and $T=0.4$ (triangles).
(b) Decay of $A(T)$ at $\rho=0.35$ (dashed) fitted to an Ising critical form $A(T)\sim|1-T/0.54|^{\beta_{\mathrm{c}}}$(solid) with $\beta_{\mathrm{c}}=0.3264$ up to $T=0.545$. We estimate $T_{\mathrm{ODT}}=0.535(5)$. (c) The average cluster size at the cmc, $\bar{n}_{\mathrm{cmc}}$ decreases as $T$ increases (dots). The line is a guide for the eye.
(d) Equilibrium FCC-cluster crystal occupancy, $n_{\mathcal{C}}$ (solid symbols), and average fluid cluster size, $\bar{n}$ (empty symbols), at $T=0.35$ (squares) and $0.40$ (triangles). Because $\bar{n}>n_\mathcal{C}$ at the cluster fluid--FCC-cluster crystal transition, most clusters in the fluid must shrink for crystallization to proceed.}
\label{fig:fig2}
\end{figure}

 \begin{figure}
\includegraphics[width=0.95\columnwidth]{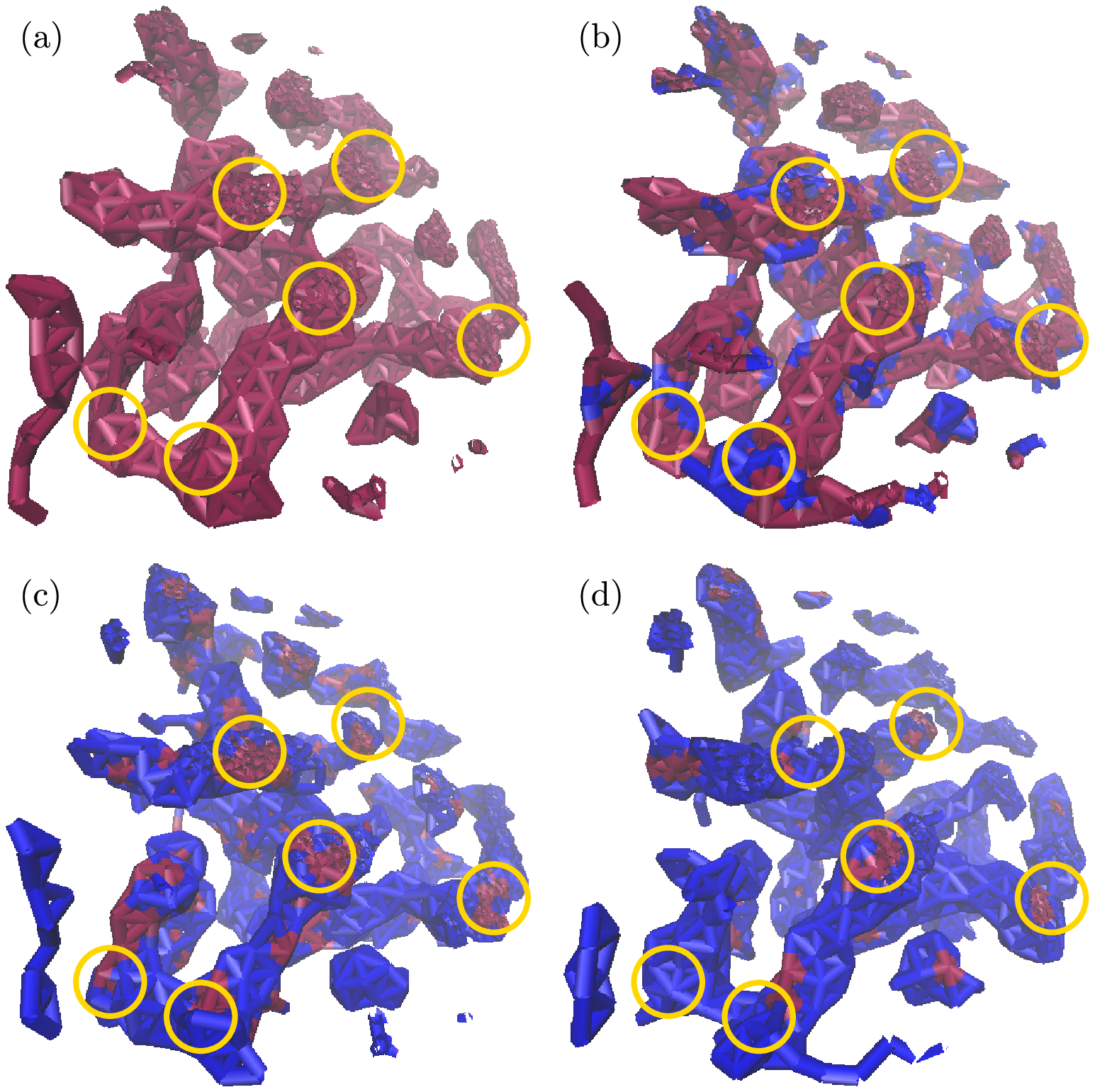}
\caption{Sample configuration slices after (a) $0$, (b) $200$, (c) $1600$, and (d) $3200$ MC sweeps of an initially equilibrated simulation with $N=2000$ at $T=0.45$ and $\rho=0.2$. The distinction between particles that remain within $0.5\sigma$ of their initial conditions (red) and those that have moved beyond it (blue) illustrates the network dynamics. Surface particles decorrelate more rapidly than core particles, and network nodes (yellow circles) reorganize more slowly than network edges.}\label{fig:fig41}
\end{figure}

\section{Simulations}
The square-well-linear (SWL) model we study here has a schematic interaction form that can be smoothly transformed into that of diblock copolymers and other microphase-forming models.
Its radial pair interaction $u(r)=u_{\mathrm{HS}}(r)+u_{\mathrm{SALR}}(r)$ includes hard-sphere volume exclusion $u_{\mathrm{HS}}(r)$ at
the particle diameter $\sigma$, which sets the unit of length, as well as a SALR contribution
\begin{equation}\label{eq:swl}
u_{\mathrm{SALR}}(r)=\left\{ \begin{array}{ccc}
  -\varepsilon &,& r<\lambda\sigma \\
  \xi\varepsilon(\kappa-r/\sigma) &,& \lambda\sigma<r<\kappa\sigma \\
  0 &,& r>\kappa\sigma\\
\end{array}\right..
\end{equation}
The square-well attraction strength $\varepsilon$, which sets the unit of energy, is felt up to
$\lambda\sigma$; beyond that point repulsion of strength $\xi\varepsilon$ takes over and decays linearly.  Note that choosing $\xi=0.05$ places the system well above the Lifshitz point, $\xi_\mathrm{L}=0.025(5)$, for the prototypical values $\lambda=1.5$ and $\kappa=4$ used here~\cite{Pini2000,Diehl2000,Seul1995}~~(see Appendix Sec.~\ref{section:TI}).
We simulate systems containing between $N=800$ and $8000$ particles under periodic boundary conditions at fixed temperature $T=1/\beta$ (the Boltzmann constant is set to unity), fixing either pressure $p$ or volume $V$ (and thus number density $\rho=N/V$).
For each state point, we perform between $10^5$ and $10^6$ Monte Carlo
(MC) sweeps, 
which include non-standard MC moves~\cite{Schultz2011,Chen2000,Frenkel2002,Whitelam2007}, in order to obtain equilibrium configurations of the different phases studied~(see Appendix Sec.~\ref{section:simulationDetails}).
The results presented here have been first equilibrated and then averaged over simulations at least five times longer than the structural relaxation time.

Obtaining equilibrium information about microscopic SALR particle-based models requires going beyond the common free energy techniques used for simulating gas, liquids and crystals~\cite{Frenkel2002}, because
these techniques fail to account for the fluctuating occupancy of periodic microphase features~\cite{Zhang2012}.
The problem is similar to that encountered in multiple-occupancy crystals~\cite{Mladek2007} and lattices with vacancies~\cite{Swope1992}.
We thus consider an expanded differential form for the Helmholtz free energy per particle~\cite{Mladek2007}
\begin{equation}
  \ud f_{\mathrm{c}} = -s\ud T-p \ud (1/\rho)+\mu_{\mathrm{c}}\ud n_{\mathrm{c}},
\end{equation}
in which the standard thermodynamic contributions, including
the entropy per particle $s$, are complemented with a field
$\mu_{\mathrm{c}}$ that is conjugate to the microphase occupancy
$n_{\mathrm{c}}$~(see Appendix Sec.~\ref{section:OccupancyAndReferenceField}). (This last quantity is generally proportional to the number of particles per period, but, for convenience, its specific definition here depends on the phase symmetry, e.g., area density $\varrho_\ell$ for lamellae of periodicity $\ell$, line density for cylinders and average cluster size $n_\mathcal{C}$ for cluster crystals.) Because in the
thermodynamic $N\rightarrow\infty$ limit $\mu_{\mathrm{c}}$ must vanish at equilibrium, optimal finite-size estimates
have $\mu_{\mathrm{c}}=0$. Standard simulation schemes cannot, however, directly
minimize this function because of the incommensurability between the
mesoscale patterns and the simulation box in finite size systems~\cite{Mladek2007,Wilding2011}. Hence, we first obtain the constrained free
energy, $f_\mathrm{c}$, of a given microphase morphology at a given ($T,\rho$) state
point and fixed $n_{\mathrm{c}}$ through a two-step thermodynamic
integration (TI) scheme: (i) from an ideal gas to a liquid of hard spheres under a modulated field; (ii) from this last state to SWL particles without field. The resulting constrained free energy is then optimized with respect to $n_{\mathrm{c}}$ (Fig.~\ref{fig:fig1})~(see Appendix Sec.~\ref{section:TI}).

\section{Phase Diagrams}
The common tangent construction is used to obtain the coexistence boundaries between different phases and thus the overall phase diagram (Fig.~\ref{fig:fig4}). As expected, at high $T$ the system is disordered, 
while at low $T$
equilibrium microphases form. Four different ordered
microphase morphologies are identified for $\rho\lesssim 0.45$:
face-centered cubic (FCC) cluster crystal, cylindrical, double gyroid
and lamellar phases (see depictions Fig.~\ref{fig:fig4} and symmetry details in Appendix Sec.~\ref{section:OccupancyAndReferenceField}.
Although a Landau functional calculation for simple microphase formers suggests that a body-centered cubic (BCC) cluster crystal phase might also form~\cite{Ciach2008,Ciach2013}, we found this structure to be only metastable in our system. The absence of the BCC symmetry suggests that the effective repulsion between clusters is
harsher than $1/r^8$~\cite{Agrawal1995,Zhang2011, Zhang2010b}, which a Hamaker-like calculation confirms~(see Appendix Sec.~\ref{section:ucc}). This morphology thus appears to be more sensitive than others to the form of the interaction potential. The other microphase morphologies considered, i.e., O70~\cite{Tyler2005}, P--surface~\cite{Gozdz1996}, ordered bicontinuous double diamond~\cite{Winey1992} and
perforated lamellae~\cite{Matsen1996},
were all found to be unstable within the regime studied.

The highest temperature at
which periodic microphases melt is the weakly first-order, order-disorder transition (ODT)~\cite{Bates1990,Seul1995}. This transition replaces the second-order gas-liquid critical point for systems beyond the Lifshitz point, i.e., for $\xi>\xi_{\mathrm{L}}$~\cite{Brazovskii1975}~((see Appendix Sec.~\ref{section:lifshitz})). Melting of the periodic lamellae at $T_{\mathrm{ODT}}$  is monitored by the decay of the order parameter
\begin{equation}
A(T)=\frac{1}{N}S(k^*;T),
\end{equation}
where $k^*$ is the low-$k$ maximum of the structure factor,
$S(k;T)$. In our model, this transition occurs roughly halfway through the lamellar regime, at $\rho\approx0.35$. Because $\varrho_\ell$, and thus $\ell$, is fairly independent of temperature in this regime (Fig.~\ref{fig:fig2}(a)), we use the $T=0.3$ value of $k^*\approx 2\pi/\ell$ to study the decay of $A(T)$. Simulation results indicate that although away from the transition the order
parameter behaves nearly critically, $A(T)$ vanishes discontinuously
at $T_{\mathrm{ODT}}=0.535(5)$ (Fig.~\ref{fig:fig2}(b)). Mechanistically, upon going through the transition lamellae become increasingly flexible, giving rise to a percolated network, as observed in diblock copolymers~\cite{Portmann2006}. 

At low temperatures, clusters form upon increasing density even before the onset of periodic microphase ordering, as reported in prior simulations and experiments~\cite{Stradner2004,Sciortino2004}.
The fluid equation of state allows us to locate the
onset of clustering, which is akin to determining the critical micelle concentration (cmc) in a surfactant system~(see Appendix Sec.~\ref{section:cmc}). Increasing $T$ along this line
decreases the average cluster size $\bar{n}_{\mathrm{cmc}}$ (Fig.~\ref{fig:fig2}(c)), and the last hints of a cmc vanish
around $T=0.72(1)$~(see Appendix Sec.~\ref{section:cmc}). Even within the fluid of clusters, the intra-cluster cohesion is relatively weak, resulting in the clusters' \emph{internal} structure to also be fluid-like. This behavior contrasts with the crystallites observed in systems with shorter attraction ranges~\cite{deCandia2006}, but lowering temperature may also lead to internally-ordered clusters in this system. In spite of their internal fluidity, the clusters are not generally spherical, and their asphericity increases with $\rho$~(see Appendix Sec.~\ref{section:clusterSize}). For $T\gtrsim0.45$, they even become wormlike and eventually percolate~\cite{Stauffer1994}~(see Appendix Sec.~\ref{section:clusterSize}), which gels the system (see below) \emph{before} the first-order transition into the periodic microphase regime is reached. This behavior is similar to that observed in Refs.~\cite{Campbell2005,Sciortino2004}, but contrasts with that of Ref.~\cite{deCandia2006}, where the percolating network was instead associated with incompletely ordered cylinder or lamellar phases. 
For $T\lesssim 0.45$, by contrast, cluster elongation is preempted by the microphase regime.  Although this last transition is reminiscent of the crystallization of purely repulsive particles, the clusters in the fluid phase are larger (Fig.~\ref{fig:fig2}(d)) and display a much wider range of sizes and morphologies than those in the FCC-cluster crystal~\cite{Mladek2007}. The FCC-cluster crystal assembly is thus expected to be more intricate than simple nucleation and growth. 

We finally consider the percolated regime observed at temperatures above the periodic microphase regime. At $T\gg T_{\mathrm{ODT}}$ the system behaves like a regular fluid, but as $T$ approaches $T_{\mathrm{ODT}}$ the structural relaxation grows increasingly complex, even under the strongly non-local MC sampling we use here (Fig.~\ref{fig:fig41}): (i) particles at the surface are lot more mobile than those in the core~\cite{Wu2002,Stevenson2008}, and (ii) edges of the network reorganize much faster than its nodes.
Being in equilibrium, the system does not age, but MC sampling is nonetheless arduous~(see Appendix Sec.~\ref{sec:SimulationDetails}), making the system gel-like. This multi-timescale dynamics, in particular the sluggish relaxation of network nodes, may contribute to the difficulty of assembling microphases in colloids~\cite{Toledano2009,Liu2005}.  Note that other mechanisms slowing down the dynamics could also emerge below $T_{\mathrm{ODT}}$, including competing microphase morphologies~\cite{Zhang2006,deCandia2011,DelGado2010} and spinodal-like arrest~\cite{deCandia2006,Tarzia2007,Charbonneau2007}, but a systematic study of these non-equilibrium effects is left for future consideration.

\section{Conclusion}
We have developed a TI-based simulation method for solving the phase diagram of arbitrary
continuous-space microphase-forming models. Our solution of
the prototypical SWL model
presents the periodic microphase sequence --
cluster crystal, cylindrical, double gyroid and lamellar phases --
of systems described by a comparable Landau functional~\cite{Ciach2013},
Our search for ordered phases, however, was not exhaustive, hence other stable morphologies are possible.
More importantly, we have clarified the thermodynamic interplay between fluids of spherical and wormlike clusters, the equilibrium percolating fluid (gel-like), and periodic microphases. This distinction is essential for separating equilibrium from non-equilibrium effects in the dynamical arrest of microphase formers~\cite{deCandia2006,Charbonneau2007,Toledano2009,Ciach2013}. It is also essential for guiding experiments with SALR-like colloidal interactions, whose precise form can vary with system density~\cite{Klix2013}. Indeed, colloidal experiments have thus far only identified equilibrium cluster fluids and gels~\cite{Stradner2004,Sciortino2004,Campbell2005,Klix2010,Zhang2012b}. Whether the challenge of assembling ordered microphases in colloids could be surmounted by tuning the properties of these disordered regimes or by identifying alternate assembly pathways remains, however, an open question.

\begin{acknowledgments}
We acknowledge many stimulating discussions about this project over the years, in particular with A.~Ciach, E.~Del Gado, D.~Frenkel, P.~Royall and S.~Yaida. We acknowledge support from the National Science Foundation Grant no. NSF DMR-1055586.
\end{acknowledgments}

\appendix
\begin{widetext}

The following material contains additional methodological details about the study of the periodic microphase and cluster fluid and percolated regimes as well as a description of the Monte Carlo sampling scheme.

\section{Periodic Microphases}
\label{sec:PeriodicMicrophases}
High free-energy barriers separate periodic microphases with different morphologies, making direct equilibration unachievable in that regime. We instead prepare the system into a given morphology and calculate its free energy at a given state point. We then compare the free energy of the different phases in order to identify the equilibrium structure. General arguments suggest that particles with isotropic SALR interactions form periodic cluster crystal,
cylindrical, double gyroid and lamellar phases~\cite{Leibler1980}. We
thus consider these morphologies as well as less common ones,
including O70~\cite{Tyler2005}, perforated lamellar~\cite{Matsen1996}, ordered
bicontinuous double--diamond~\cite{Matsen1996} and simple gyroid~\cite{Scherer2013}.

\subsection{Lattice Occupancy and Reference Field}
\label{section:OccupancyAndReferenceField}
In order to obtain the equilibrium free energy of a given morphology at a given state point, we define a specific measure of lattice occupancy and a reference modulated field (Table~\ref{tab:occ}). The former is a parameter with respect to which the free energy is optimized, which is necessary because the equilibrium periodicity of finite-size systems may be incommensurate with the prepared simulation box. The latter is chosen so as to obtain a continuous thermodynamic integration path from a reference ideal system whose free energy is known. The field must thus break the same symmetry as the periodic phase, while remaining simple to formulate.
When possible, we use combinations of trigonometric functions with a wavevector $k=2\pi/\ell$ to control the periodicity $\ell$ of the field. For example, for the lamellar phase, which is periodic in one dimension, we use a simple sinusoidal function, for which lattice occupancy is measured by taking the particle number density per unit area of lamella,
$\varrho_\ell$. Note that $\varrho_\ell \equiv N_T/A = \rho \ell$, where $N_T$ is the total number of particles in one period, and $A$ is the area of a lamella.
The double gyroid morphology, which is a bicontinuous phase that emerges from a minimal surface problem, is, however, more intricate to describe. In that case, the field that couples to the particle distribution is a minimal surface function.

\begin{table}
\caption{Measures of occupancy and reference field for the relevant microphase morphologies}
\label{tab:occ}
\begin{tabular}{p{1.83cm} p{3.2cm} c c p {8.1cm}}
  \hline   \hline
  Morphology &         Measure of occupancy & $n_{\mathrm{c}}$ symbol & Typical range & Reference field, $\mathcal{F}(\br)$\\
  \hline   \hline
  Lamellar &          Average area density per layer  & $\varrho_\ell$ & $1-2.5$ & $\mathcal{F}_0\cos (kz)$ \\   \hline{}
  Cylindrical &     Average line density per cylinder & & $6-12$ & $\begin{aligned}\mathcal{F}_0 &\cos (kx)\cos\left[k
    \left(
      \frac{x}{2}+\frac{\sqrt{3}y}{2}
    \right)\right] \cos\left[k
    \left(
      \frac{x}{2}-\frac{\sqrt{3}y}{2}
    \right)\right]\end{aligned}$\\   \hline
  FCC-cluster crystal    & Average number of particles per cluster &   $n_\mathcal{C}$    & $25-40$ & $\mathcal{F}_0 \cos(kx)\cos(ky)\cos(kz)$\\   \hline{}
    BCC-cluster crystal &   Average number of particles per cluster & $n_\mathcal{C}$       & $25-40$ & $\begin{aligned}&\mathcal{F}_0 \left|\cos[k(x+y)]\cos[k(x-y)]\cos[k(y+z)]\cos[k(y-z)]\right.
    \\ &\left.\ \cos[k(z+x)]\cos[k(z-x)]\right|^{1/2}
    \end{aligned}$\\     \hline

  Double Gyroid &           Average number of particles
                                            per unit cell. & & $600-800$ & $\begin{aligned}-\mathcal{F}_0 \sin \left[g(x,y,z)^2 \right]&,
                                            \\  \mathrm{where\quad} g(x,y,z)=&\cos(kx)\sin(ky)+\cos(ky)\sin(kz) \\&+\cos(kz)\sin(kx)\end{aligned} $ \\   \hline
  \hline{}
\end{tabular}
\end{table}

\subsection{Two-Step Thermodynamic Integration}
\label{section:TI}
The free energy of the periodic microphases is determined by a two-step thermodynamic integration (TI) scheme. The reference state consists of low-density hard spheres under an external modulated field (HSF), whose free energy is
\begin{equation}
f_{\mathrm{c},0}^{\mathrm{HSF}}=f^{\mathrm{id}}(\rho)+\int_Ve^{-\beta B(\br)}\ud\br,
\end{equation}
where $f^{\mathrm{id}}(\rho)=\ln(\rho\Lambda^3)/\beta$
is the free energy of an ideal gas of number density $\rho$. The thermal de Broglie wave length, $\Lambda$, is set to unity without loss of generality.

The first TI step brings the system density from $\rho=0$ up to the targeted density as
\begin{equation}
f_{\mathrm{c}}^{\mathrm{HSF}}(\rho)=f_{\mathrm{c},0}^{\mathrm{HSF}}+\int_0^{\rho}\frac{p(\rho')-\rho'
  k_B T}{\rho'^2}\ud\rho',
\end{equation}
while keeping the field fixed.
At low densities the first two virial coefficients are used to facilitate numerical integration (Sect.~\ref{section:virial}). Note that for the lamellar and cylindrical phases, a virtual harmonic spring
\begin{equation}
U_z = k_z(L_z-L_{z_{0}})
\end{equation}
with stiffness $k_z$ is also applied to the periodic directions of the
box. This spring ensures that the final configuration preserves the targeted lattice occupancy. Because the integrand is here independent of the system energy, the integration is unaffected by this constraint.

The second TI step, which brings HSF to fully-interacting SWL particles, follows the linear alchemical transformation,
$U_\delta(\textbf{r}^N)=-(1-\delta)\sum_{i=1}^{N}B(\textbf{r}_{i})+\delta\sum_{i>j=1}^N u_{\mathrm{SALR}}(r_{ij})$, which gives
\begin{equation}
  f_\mathrm{c}(\rho)=f_\mathrm{c}^{\mathrm{HSF}}(\rho) + \frac{1}{N}\int_{\delta=0}^{\delta=1}\ud\delta
\left<\frac{\partial U_\delta}{\partial \delta} \right>_\delta,
\end{equation}
for
A Gauss--Lobatto quadrature with $20$ points is
used to minimize the numerical integration error.

Summing the various
contributions gives the free energy of a system constrained to a given morphology and lattice occupancy\begin{equation}
  f_\mathrm{c}(\rho)=f^{\mathrm{id}}(\rho)+\int_Ve^{-\beta B(\br)}\ud\br+\int_0^{\rho}\frac{p(\rho')-\rho'
  k_B T}{\rho'^2}\ud\rho +\frac{1}{N}\int_{\delta=0}^{\delta=1}\ud\delta
\left<\frac{\partial U_\delta}{\partial \delta} \right>_\delta.
\end{equation}

\subsection{HSF Virial Coefficients}
\label{section:virial}
The first two virial coefficients for HSF, $B_2$ and $B_3$, correspond to the intercept and the initial slope, respectively, of the integrand for the first TI step. Because simulation
results suffer from error magnification at low $\rho$, these two quantities are particularly useful.
$B_2$ is evaluated by integrating over particle positions $\br_i$ under the external field $\mathcal{F}(\br)$,
\begin{equation}
B_2=\int_{\br_1}\int_{\br_2}\ud\br_1\ud\br_2 f_{12}
\frac{e^{\beta\mathcal{F} (\br_1)}e^{\beta\mathcal{F}(\br_2)}}{\left(\int_{\br}\ud\br e^{\beta\mathcal{F}(\br)}\right)^2 },
\end{equation}
where for hard spheres the Mayer function is simply $f_{ij}=-\Theta(\sigma-|\br_1-\br_2|)$, with $\Theta(r)$ the Heaviside Theta function. The expression for lamellae can be simplified as
\begin{equation}
  B_2=\pi\int_0^\sigma\ud r \int_0^\pi\ud \theta \int_0^{L_z}\ud z \frac{e^{-\beta\mathcal{F}_0\sin\left[k_z\left(r\cos\theta+z\right)\right]+\sin {k_zz}}}{L_z\left(\int_0^{L_z}\ud z e^{-\beta\mathcal{F}_0\sin{k_zz}}\right)^2},
\end{equation}
but for the other fields the full expression is numerically evaluated by Monte Carlo integration. We can similarly obtain and evaluate
\begin{equation}
  B_3=\int_{\br_1}\int_{\br_2}\int_{\br_3}\ud\br_1\ud\br_2\ud\br_3 f_{12}f_{13}f_{23}
\frac{e^{\beta\mathcal{F} (\br_1)}e^{\beta\mathcal{F}(\br_2)}e^{\beta\mathcal{F}(\br_3)}}{\left(\int_{\br}\ud\br e^{\beta\mathcal{F}(\br)}\right)^3 }.
\end{equation}
Sample results are given in Fig.~\ref{fig:TB3}.
\begin{figure}
  \includegraphics[width=0.32\textwidth]{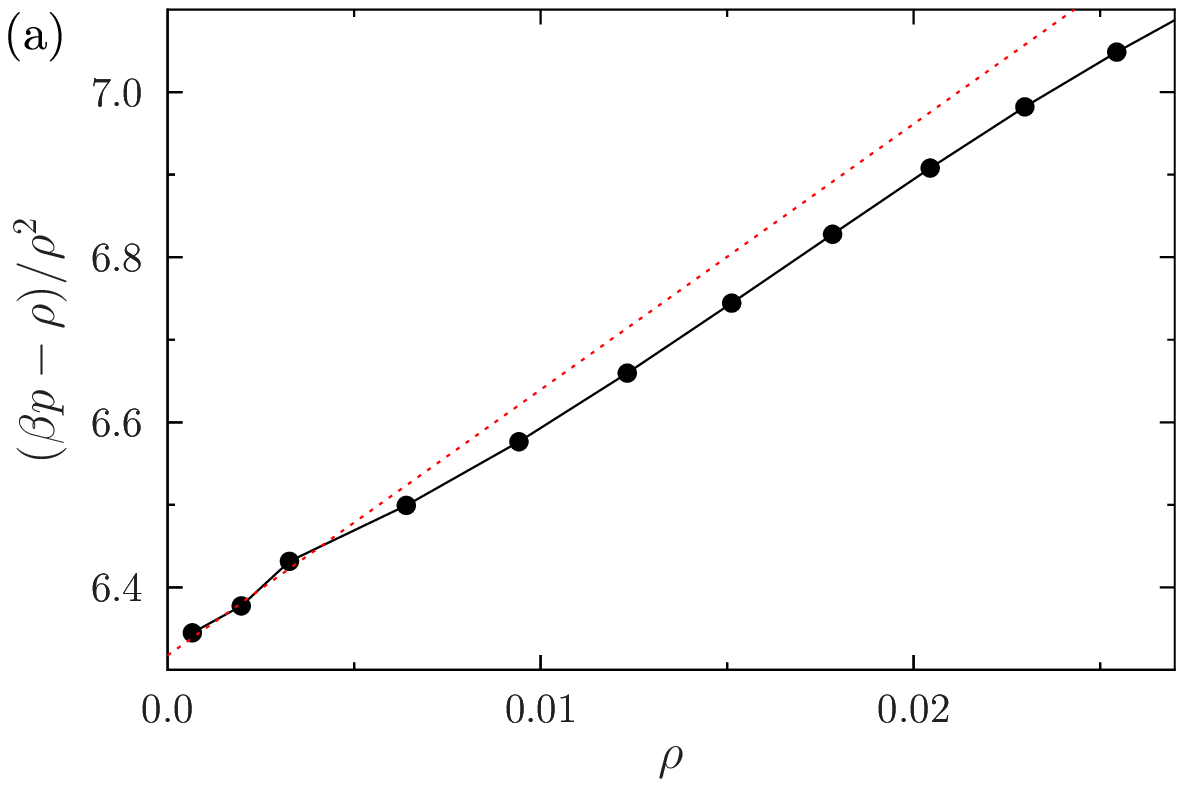}
        \includegraphics[width=0.32\textwidth]{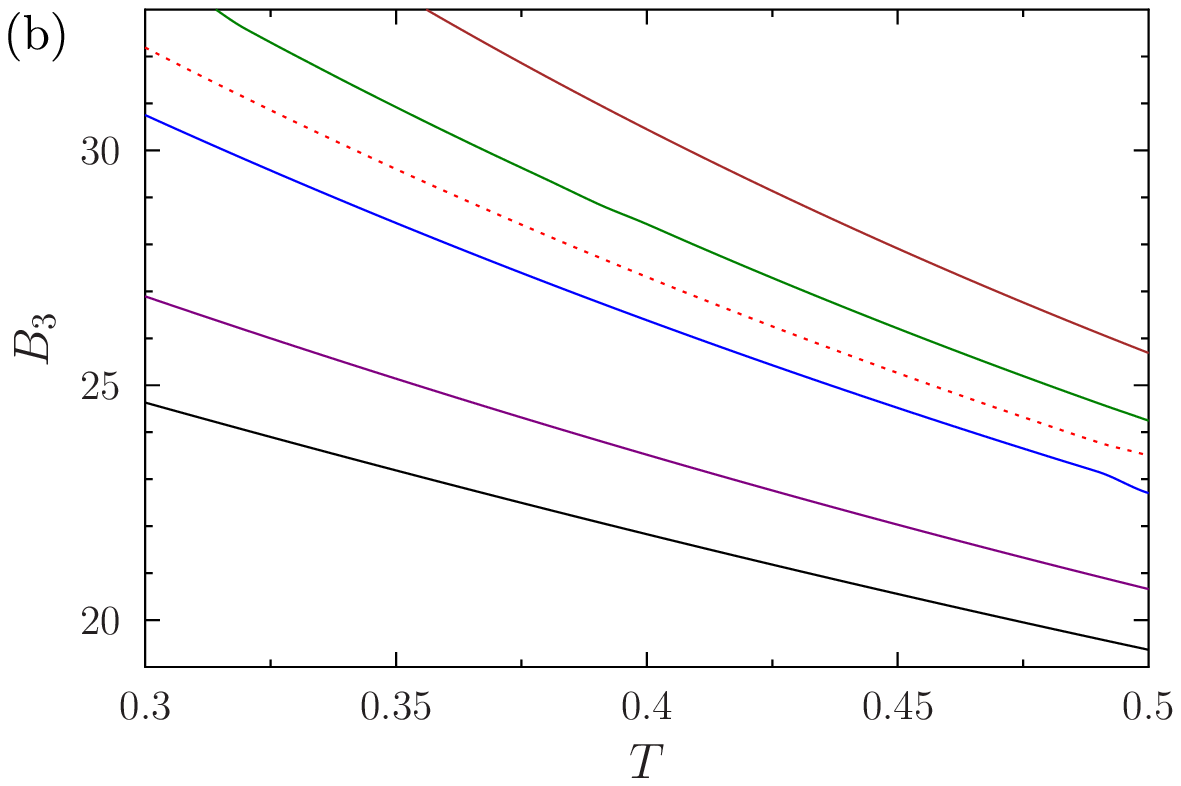}
        \includegraphics[width=0.32\textwidth]{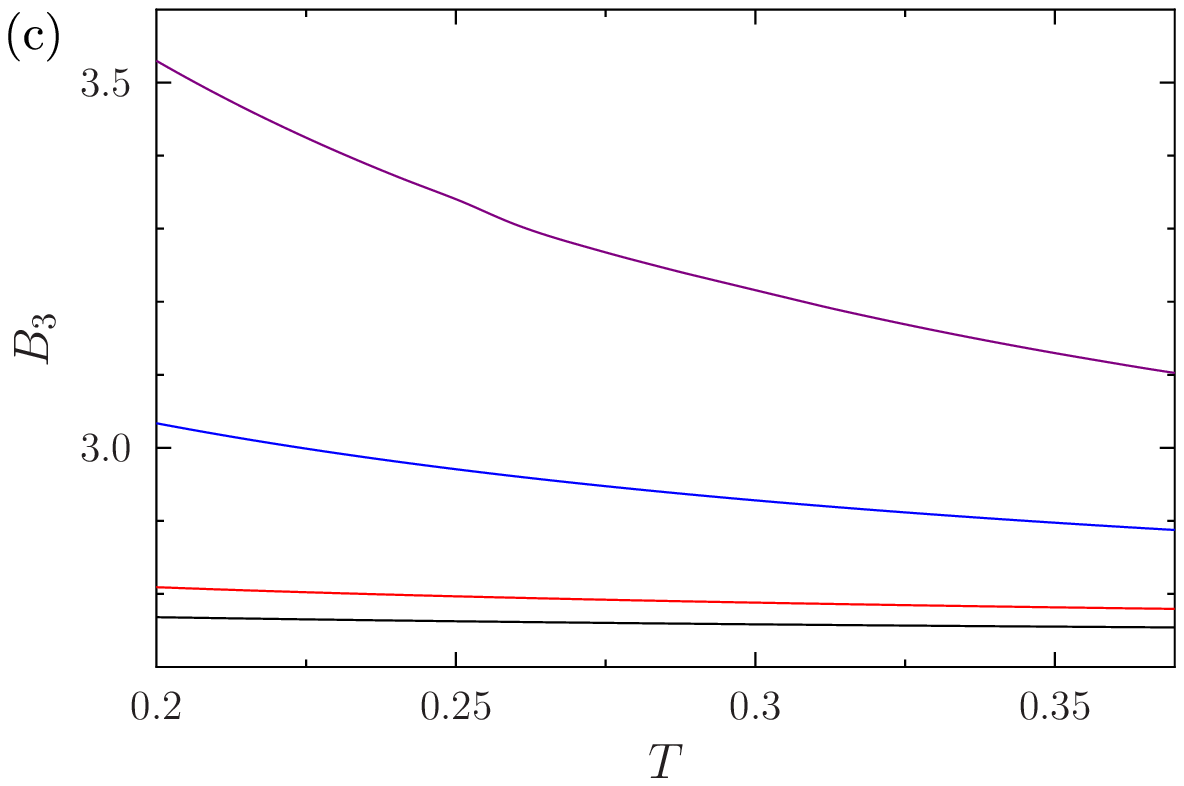}

        \caption{(a) Equation of state for hard spheres a under
          modulated-lamellar field at $T=0.3$, $\ell=5.25$ from simulations (dots) and from the virial expansion with $B_2=6.318$ and $B_3=32.19$ (dashed line). Third virial coefficient, $B_3$, for the reference field of (b) the lamellar phase for interlayer $\ell= 4.2$, $4.5$, $5$, $5.25$, $5.5$ and $6$ (from bottom to top), and (c) the FCC-cluster crystal phase for intercluster $\ell=164.1$, $76.17$, $35.36$ and $22.27$ (from top to bottom). 
          }  \label{fig:TB3}
\end{figure}

\subsection{Lifshitz point}
\label{section:lifshitz}
Microphases form in systems with SALR interactions for which the repulsion is sufficiently strong. For weak repulsion, a simple gas-liquid binodal is found in their stead. For anisotropic repulsion the transition between the two regimes produces a tricritical Lifshitz point~\cite{Diehl2000}, but for isotropic repulsion the order-disorder transition into the microphase regime is weakly first-order. Although the Lifshitz point then formally disappears~\cite{Brazovskii1975, Seul1995}, some authors have elected to keep the same vocable to describe the transition from a gas-liquid critical point to microphase formation~\cite{Pini2000}. We follow this convention.

In order to ensure that our SWL potential falls on the microphase-forming side of the Lifshitz point, we estimate its location with respect to the repulsion strength, $\xi$. Standard Gibbs ensemble Monte Carlo (GEMC) simulations with two simulation boxes, initially each with $N=512$, for $10^6$ steps locate the gas-liquid coexistence regime (Fig.~\ref{fig:potentials}).
\begin{figure}
        \includegraphics[width=0.48\textwidth]{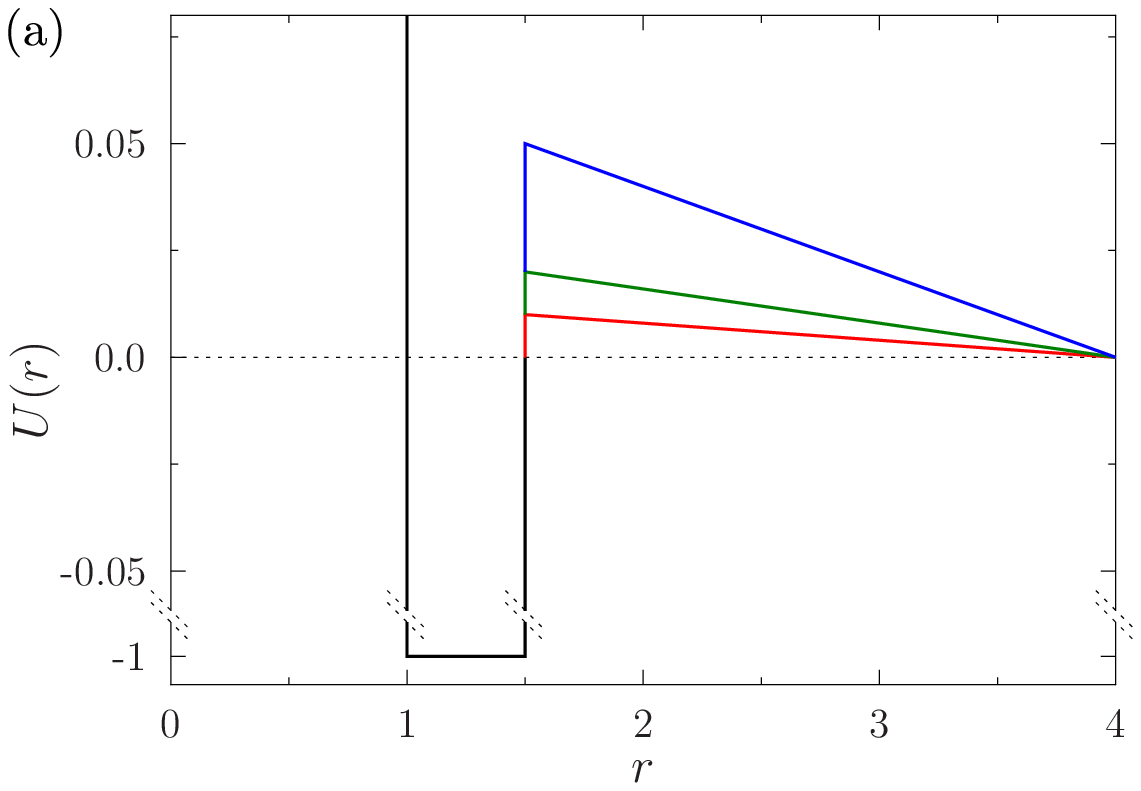}
        \includegraphics[width=0.48\textwidth]{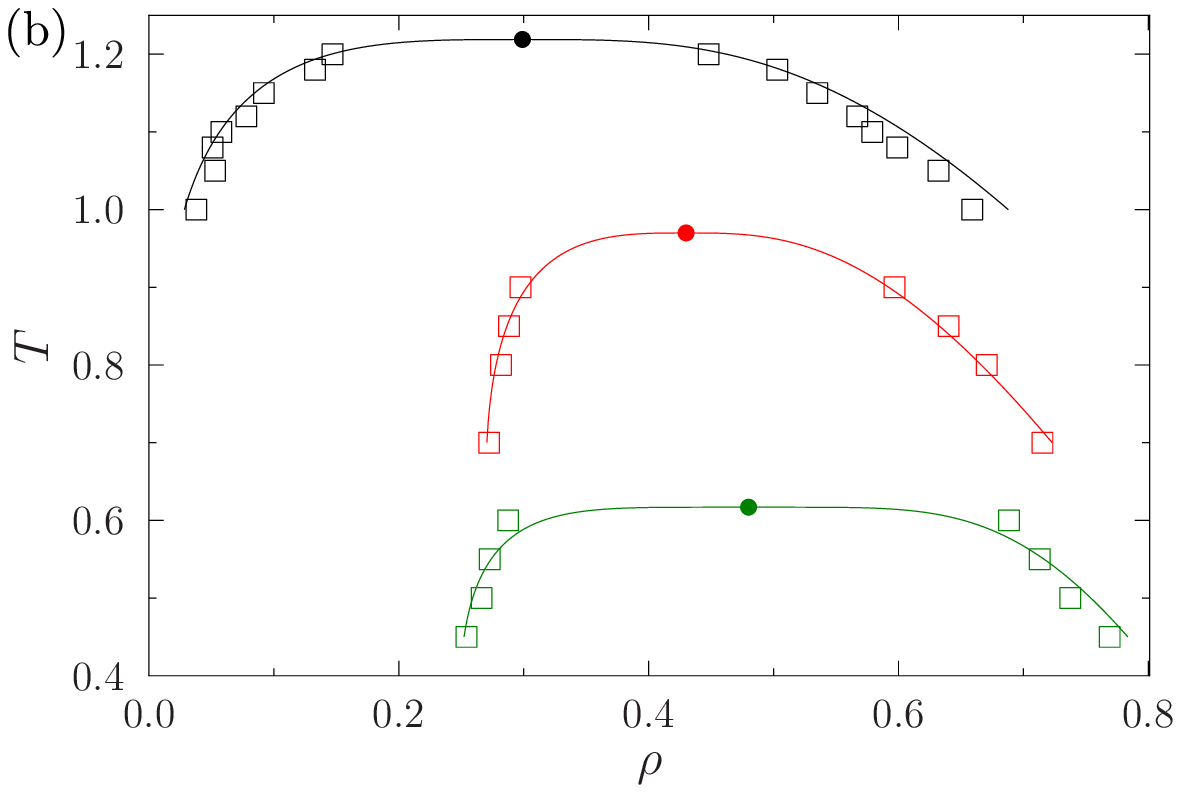}
        \caption{(a) SALR interaction potentials for $\xi=0$ (black), $\xi=0.01$ (red), 0.02 (green) and 0.05 (blue). (b) GEMC simulation results (squares) for different $\xi$ show the accompanying evolution of the critical temperature $T_\mathrm{c}$ (dots). Lines are fit to Eq.~\eqref{eq:crit}. Note that for $\xi=0$ the results are consistent with those of Ref.~\cite{Liu2005}, and that for $\xi=0.02$, the critical temperature $T_\mathrm{c}=0.61(1)$ is quite close to $T_{\mathrm{ODT}}=0.535(5)$ for $\xi=0.05$.}
        \label{fig:potentials}
\end{figure}
Fitting the coexistence results with
\begin{equation}
\label{eq:crit}
  \rho_{\pm} = \rho_{\mathrm{c}}+2C_2\left|1-\frac{T}{T_\mathrm{c}}\right|\pm \frac{1}{2}B_0\left|1-\frac{T}{T_\mathrm{c}} \right|^{\beta_{\mathrm{c}}},
\end{equation}
where $C_2$ and $B_0$ are fit parameters and $\beta_{\mathrm{c}}=0.3264$ is the three-dimensional Ising universality class critical exponent, provides an estimate of the gas-liquid critical temperature $T_{\mathrm{c}}$ and density $\rho_{\mathrm{c}}$. The results for $\xi\leq0.02$ are given in Fig.~\ref{fig:potentials}, but $\xi\gtrsim0.03$ no gas-liquid coexistence regime can be detected. We thus estimate the Lifshitz point, $\xi_{\mathrm{L}}=0.025(5)$, which indicates that that our SWL model at $\xi=0.05$ is well within the microphase-forming regime.

\section{Cluster properties}
\subsection{CMC Determination}
\label{section:cmc}
Standard approaches for determining the critical micelle concentration (cmc) implicitly rely on the existence of ideal-gas-like scaling in both the particle and the cluster fluid regimes of the equation of state. Such an ideal scenario only exists for systems in which the cluster fluid regime has a fairly broad density range of stability and for which the distribution of cluster sizes, $P(n)$, is narrow and roughly independent of particle concentration. Because neither of these conditions is met here, we develop a more general scheme for estimating the cmc.

In the spirit of the classical definition of the cmc, which situates the crossover as the point of most abrupt change in the physico-chemical properties of the system~\cite{Gelbart2012}, we take the minimum -- when it exists -- of
\begin{equation}
h(\rho)\equiv\frac{\beta p-\rho}{\rho^2}.
\end{equation}
This function measures deviations of the equation of state from the ideal gas behavior. The pronounced peak it exhibits in cluster-forming systems (see Fig.~3 of the main text) indicates a rapid change in the system properties.
In practice, we determine the cmc by first fitting the numerical results for the equation of state to
\begin{equation}\label{eq:cmcpressure}
 \beta p=\rho_\mathrm{s}+\rho_\mathrm{m}+B_{2,\mathrm{ss}}\rho_\mathrm{s}^2+B_{2,\mathrm{sm}}\rho_\mathrm{s}\rho_\mathrm{m}+B_{2,\mathrm{mm}}\rho_\mathrm{m}^2,
\end{equation}
where $\rho_{\mathrm{s}}$ and $\rho_{\mathrm{m}}$  are the monomer and
cluster number density, respectively. Note that the second virial
coefficients $B_{2,\mathrm{mm}}$ and $B_{2,\mathrm{sm}}$ are fitted to the simulation data, but we directly calculate
\begin{equation}
B_{2,\mathrm{ss}} = -\frac{2\pi}{3}\left\{-\sigma^3+\sigma^3[\lambda^3-1][e^{\beta\varepsilon}-1] + 6\int_{\lambda\sigma}^{\kappa\sigma}r^2\left[e^{-\beta\xi\varepsilon(\kappa-r/\sigma)}-1\right]\ud r\right\}.
\end{equation}

In the rest of this section, we demonstrate that the above approach reduces to the standard scheme for the ideal scenario~\cite{Israelachvili2011}. We first consider the classical derivation. For a monodisperse distribution of cluster sizes, i.e., $P(n')=\delta(n-n')$, we can write a chemical equilibrium between individual particles and clusters,
\begin{displaymath}
  n\mathrm{S}\leftrightharpoons \mathrm{S}_n,
\end{displaymath}
with the equilibrium constant
\begin{equation}
  K\equiv\frac{\rho_{\mathrm{m}}}{\rho_{\mathrm{s}}^n}.
\end{equation}
Recalling that the total concentration of material $\rho$ is conserved, i.e., $n\rho_{\mathrm{m}}+\rho_{\mathrm{s}}=\rho$, and defining $\alpha(\rho)\equiv \rho_{\mathrm{m}}/\rho$ gives
\begin{equation}\label{eq:massAction}
  K=\frac{\alpha\rho}{(1-n\alpha)^n\rho^n},
\end{equation}
where the explicit dependence of $\alpha$ on $\rho$ is dropped in order to simplify the notation. This equation can be rewritten  as
\begin{equation}
  \ln \rho = -\frac{1}{n-1}\ln K+\frac{1}{n-1}\ln\frac{\alpha}{(1-n\alpha)^n}.
\end{equation}
At the cmc, typical values for $n\alpha^*$ range from $0.01\text{--}0.1$. Because $\ln K$ is typically much larger than that, the second term can be neglected. Hence, for $n\gg1$ we get
\begin{equation}
\label{eq:cmc}
  \rho_{\mathrm{cmc}}=K^{-1/n}.
\end{equation}
This last result is the classical definition of the cmc.

Using the scheme proposed here for the ideal scenario, we write for $n\gg1$
\begin{displaymath}
  h(\rho) = \frac{\rho_{\mathrm{s}}+\rho_{\mathrm{m}}-\rho}{\rho^2}
  =\frac{\alpha(1-n)}{\rho}.
\end{displaymath}
Defining the cmc as the minimum of $h(\rho)$, we then obtain
\begin{displaymath}
\left.\frac{\partial h(\rho)}{\partial\rho}\right|_{\rho=\rho_{\mathrm{cmc}}}=0=-\frac{(1-n)\alpha^*}{\rho_{\mathrm{cmc}}^2}+\left.\frac{(1-n)}{\rho_{\mathrm{cmc}}}\left(\frac{\partial \alpha}{\partial \rho}\right)\right|_{\rho=\rho_{\mathrm{cmc}}}=0,
\end{displaymath}
where $\frac{\partial \alpha}{\partial \rho}$ can be obtained from the law of mass action in Eq.~(\ref{eq:massAction}),
\begin{displaymath}
  K=\frac{\alpha^*\rho}{(\rho-n\alpha^*\rho)^n}\approx\frac{\alpha^*\rho}{\rho^n(1-n^2\alpha^*)}.
\end{displaymath}
For $n\alpha\ll 1$, we finally obtain
\begin{equation}
  \alpha^* = \frac{K\rho^{n-1}}{1+n^2K\rho^{n-1}},
\end{equation}
and thus
\begin{equation}
  (n-1)\log \rho_{\mathrm{cmc}}+\log K + \log\frac{n^2}{n-2}=0.
\end{equation}
In the ideal scenario, we finally recover Eq.~(\ref{eq:cmc}).

\subsection{Effective cluster-cluster interaction}
\label{section:ucc}
\begin{figure}
  \raisebox{-0.5\height}{\includegraphics[width=0.32\textwidth]{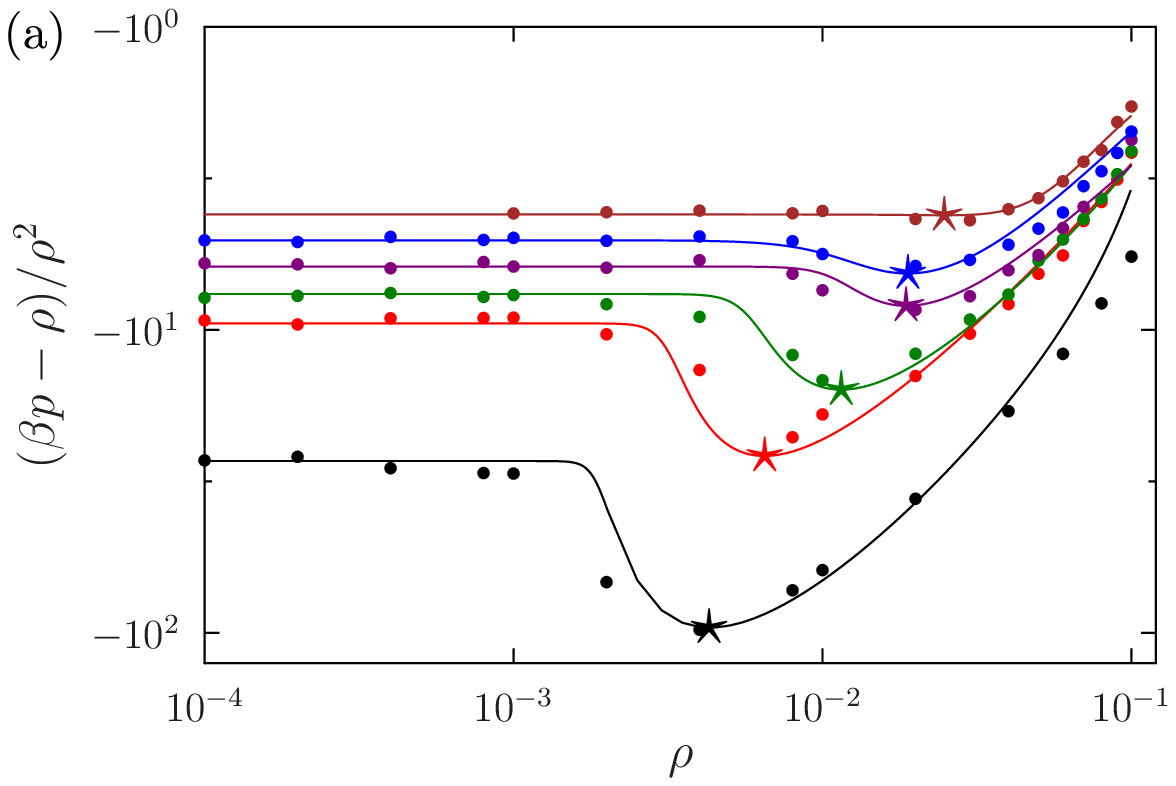}}
  \raisebox{-0.5\height}{\includegraphics[width=0.32\textwidth]{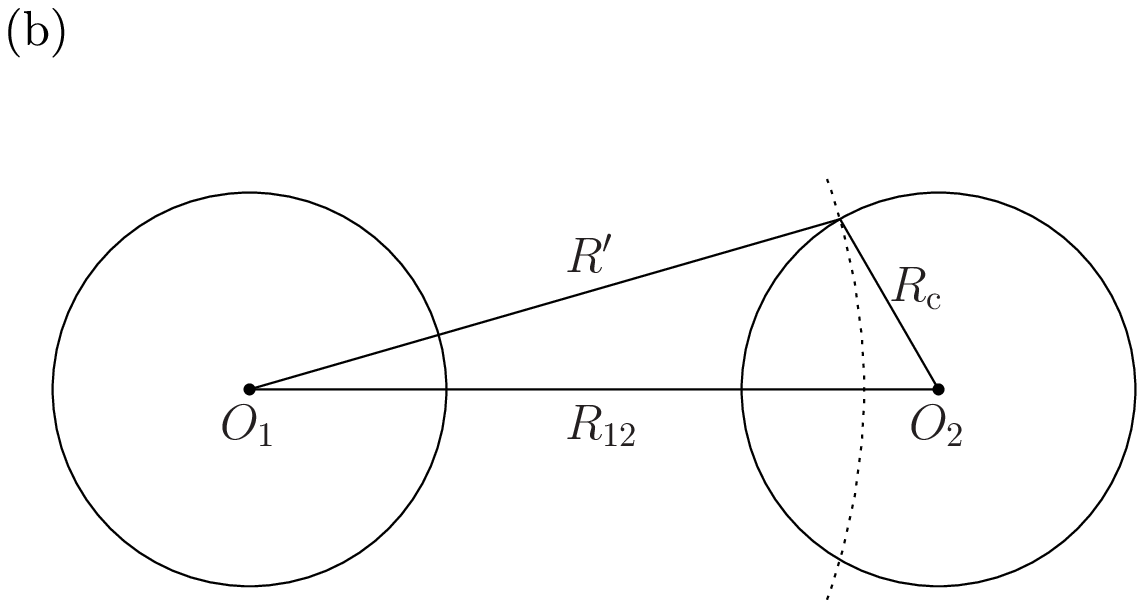}}
  \raisebox{-0.5\height}{\includegraphics[width=0.32\textwidth]{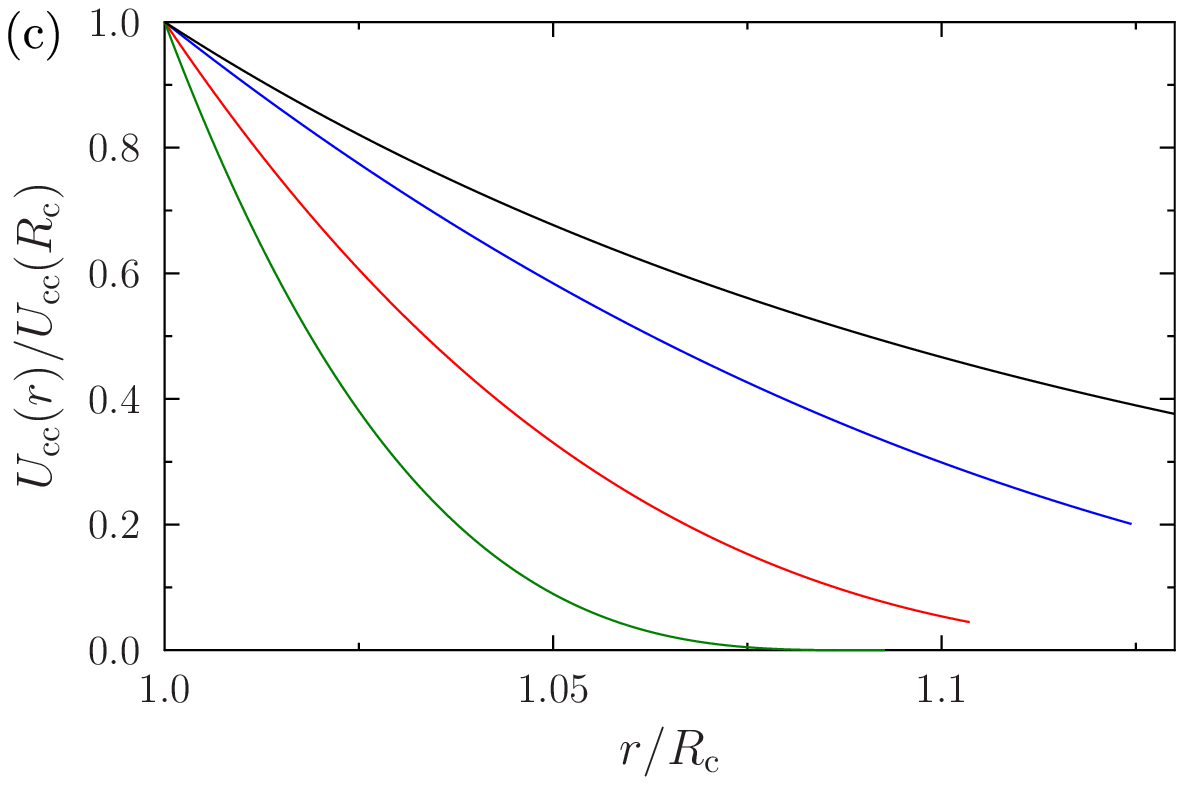}}
                \caption{(a) Fluid equations of state for $T=0.70$, $0.65$, $0.60$, $0.55$, $0.50$, $0.45$ (points,
                  from top to bottom) fitted to Eq.~\eqref{eq:cmcpressure} (lines).
                  The cmc is the point where the physical properties of the system change most rapidly, i.e., the curve minima (stars).
                  The absence of a detectable minimum above $T=0.72(1)$ signals the end of the cluster fluid regime.
                  (b) Implementation of Hamaker's integration
                  method between two homogeneous clusters. (c)
                  Comparison of the decay of the (normalized)
                  effective cluster-cluster interactions for different
                  cluster sizes, $n=10$ (blue),
                  $n=20$ (red) and
                  $n=30$ (green), using Hamaker's
                  method. Interactions between two clusters decrease
                  much more quickly than $1/r^8$ (black).
                  }
                \label{fig:ucc}
\end{figure}

By analogy with soft spheres, we expect that a sufficiently steep effective cluster-cluster interaction, $U_{\mathrm{cc}}^{\mathrm{eff}}(R')$, should result in their crystallization to a FCC-cluster crystal phase~\cite{Agrawal1995,Zhang2010b}. By assuming that
the particle density inside spherical clusters is constant and that two clusters cannot overlap, we obtain
\begin{equation}\label{eq:ucc}
  U_{\mathrm{cc}}^{\mathrm{eff}}(|\bR-\bR^{\prime}|)=
  \int \ud\br \ud\br^{\prime}P(\br-\bR) u_{\mathrm{SWL}}(\br-\br^{\prime})P(\br^{\prime}-\bR^{\prime}),
\end{equation}
where $P(\br-\bR)$ is the probability distribution function of a
particle at distance $\br$ from the center of the cluster $\bR$.
Defining the particle-cluster interaction
\begin{equation}\label{eq:uvceff}
U_{\mathrm{vc}}^{\mathrm{eff}}(|\bR-\bR^{\prime}|)=\int \ud\br^{\prime} \phi_{\mathrm{SALR}}(\bR-\br^{\prime})P(\br^{\prime}-\bR^{\prime})
\end{equation}
and using Hamaker's method~\cite{Israelachvili2011}, we get for clusters of radius $R_{\mathrm{c}}$
\begin{equation}
 U_{\mathrm{vc}}^{\mathrm{eff}}(r;R_{\mathrm{c}},\rho_{\mathrm{l}})=\rho_{\mathrm{l}}\int_{r-R_{\mathrm{c}}}^{r+R_{\mathrm{c}}}\ud R S(r,R,R_{\mathrm{c}})\phi_{\mathrm{SALR}}(r),
\end{equation}
where $S(r,R,R_{\mathrm{c}})=\frac{\pi  r \left[R_{\mathrm{c}}^2-(R-r)^2\right]}{R}$. Using the integration setup described in Fig.~\ref{fig:ucc},
a particle within cluster $O_2$ at a
distance $R'$ has an interaction energy with cluster $O_1$, $U_{\mathrm{vc}}^{\mathrm{eff}}(R')\rho_{\mathrm{l}}$.
Integrating all the particle positions in $O_2$, we thus obtain
the cluster-cluster interaction
\begin{equation}
  U_{\mathrm{cc}}^{\mathrm{eff}}(r;R_{\mathrm{c}},\rho_{\mathrm{l}})=\rho_{\mathrm{l}}\int_{R_{12}-R_{\mathrm{c}}}^{R_{12}+R_{\mathrm{c}}} \ud R^{\prime}S(R_{12},R',R_{\mathrm{c}})U_{\mathrm{vc}}^{\mathrm{eff}}(R'),
\end{equation}
Results for different cluster sizes $n=\rho_{\mathrm{l}}V_{\mathrm{c}}$(Fig.~\ref{fig:ucc}) suggest that the effective cluster-cluster interaction decays faster than $\sim1/r^8$ and thus falls within the stability regime for the FCC crystal phase.

\subsection{Cluster Size Distribution}
\label{section:clusterSize}
\begin{figure}
  \includegraphics[width=0.32\textwidth]{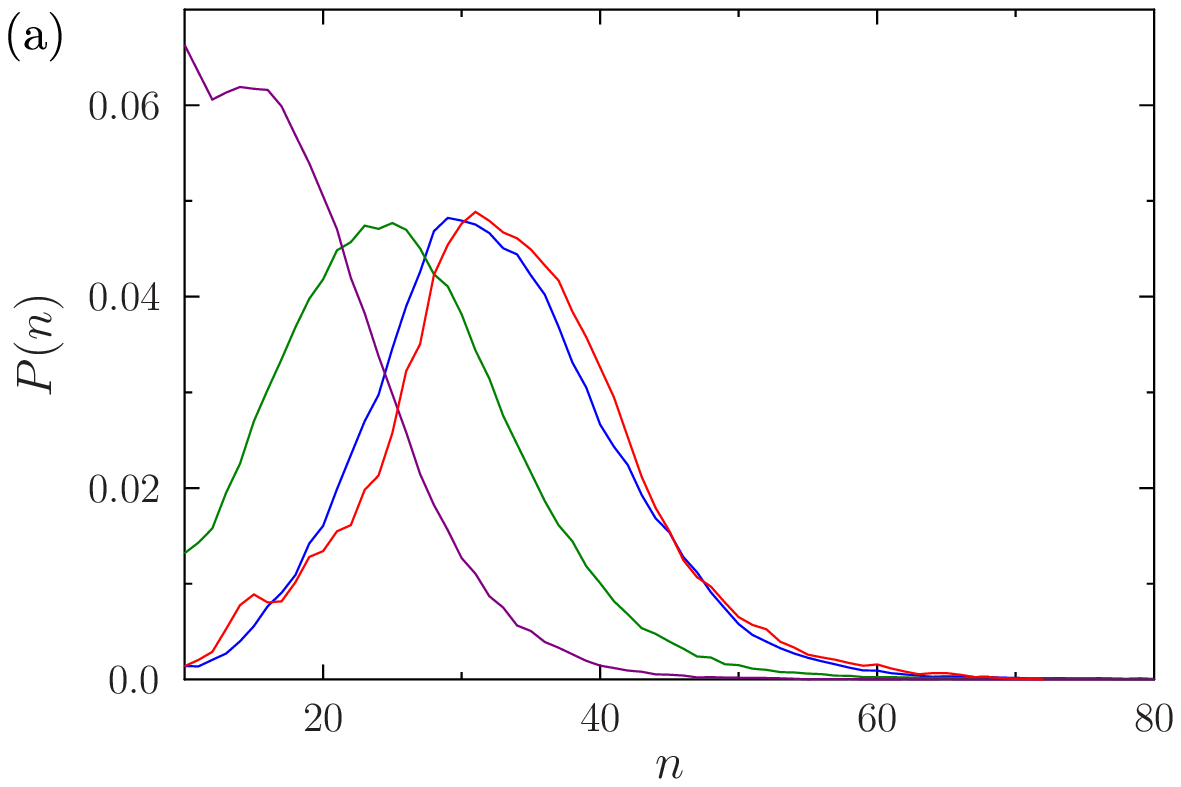}
  \includegraphics[width=0.32\textwidth]{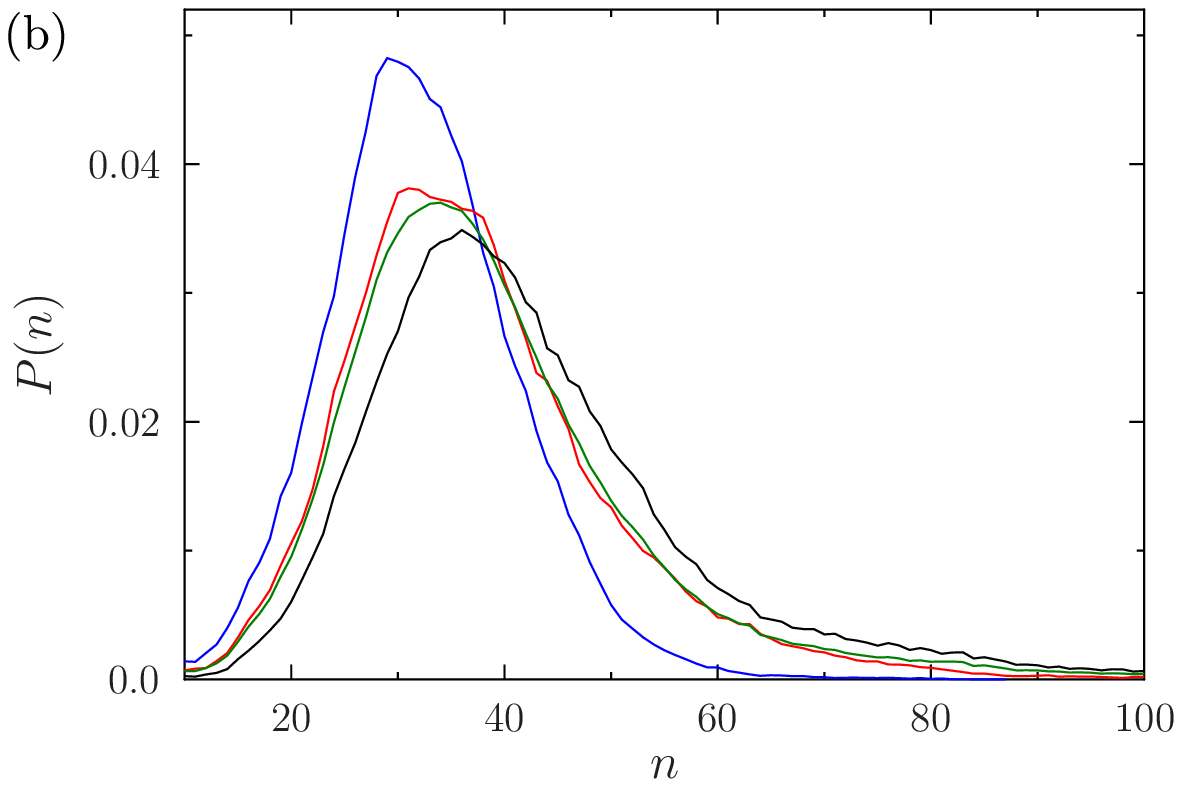}
  \includegraphics[width=0.32\textwidth]{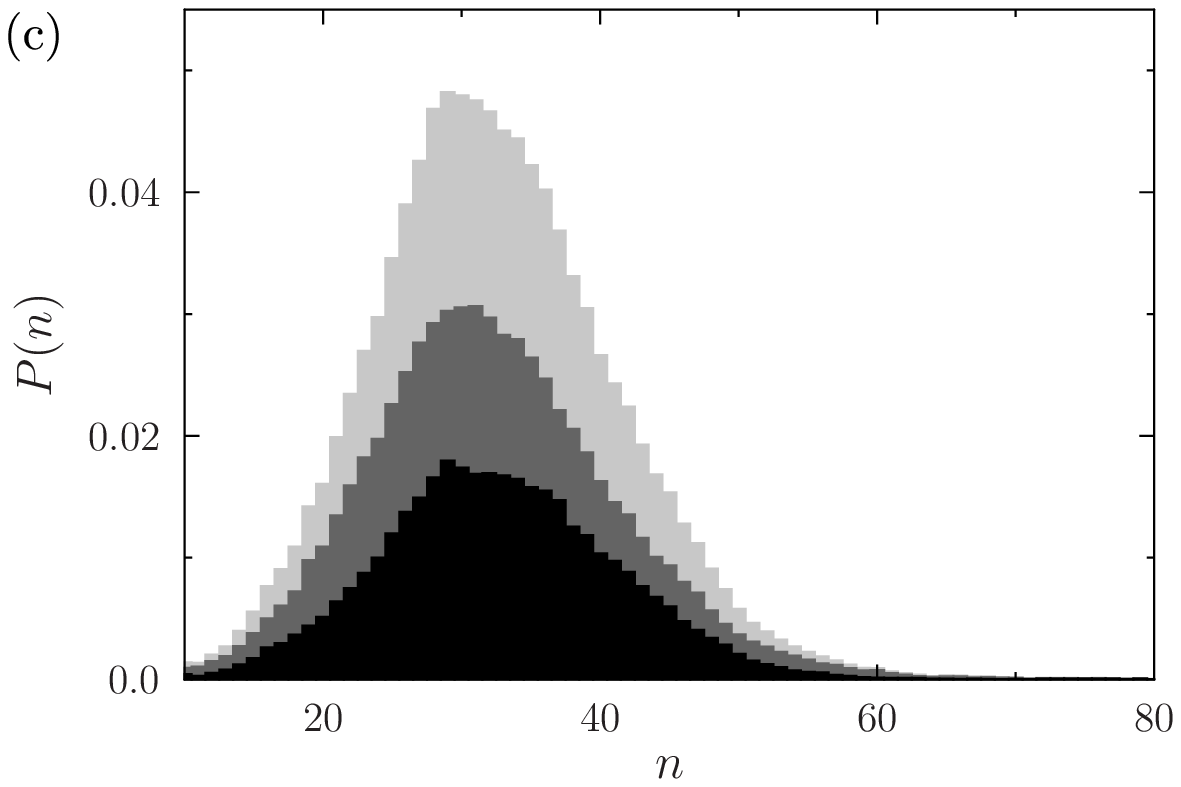}
  \caption{(a) Probability distribution function (PDF) of cluster size, $P(n)$, at the cmc for
    $T=0.35$ (red), $T=0.4$ (blue), $T=0.45$ (purple) and
    $T=0.5$ (green). The distribution steadily shifts towards smaller cluster sizes as $T$ increases. (b) PDF of cluster sizes at $T=0.4$ for
    $\rho=0.01$ (blue), $\rho=0.03$ (red), $\rho=0.05$ (green) and
    $\rho=0.07$ (black).  The distribution becomes increasingly broad with increasing density.
    (c) Total (gray), non-spherical (dark gray) and spherical (black) PDF of cluster sizes at $\rho=0.01$ and $T=0.4$, using $\zeta_{\mathrm{th}}=0.2$.
    Recall that the average of $n$ is $\bar{n}$.}\label{fig:clusterDistribution}
\end{figure}
A broad range cluster shapes and sizes is observed in the cluster
fluid regime. In Fig. 2 of the main text only average values are reported, but in Fig.~\ref{fig:clusterDistribution}, we consider the full cluster size distribution. In order to separate spherical from non-spherical clusters, we also define the anisotropy parameter $\zeta$
\begin{equation}
  \zeta = \frac{(I_1-I_2)^2+(I_1-I_3)^2+(I_2-I_3)^2}{2(I_1^2+I_2^2+I_3^2)}
\end{equation}
from the eigenvalues $I_{i}$ of the inertia tensor of each cluster. A perfectly spherical object would have $\zeta=0$.
Breaking down the probability distribution between spherical and
non-spherical clusters using an arbitrary threshold
$\zeta_{\mathrm{th}}$ suggests that a continuum of deformations from
sphericity is observed (Fig.~\ref{fig:clusterDistribution}). 
Note that this behavior is rather different from what is observed in
the FCC-cluster phase, where not only the average cluster size, but also the size and shape variances
are all reduced. 

\subsection{Percolation}
\label{section:percolation}
\begin{figure}
        \includegraphics[width=0.48\textwidth]{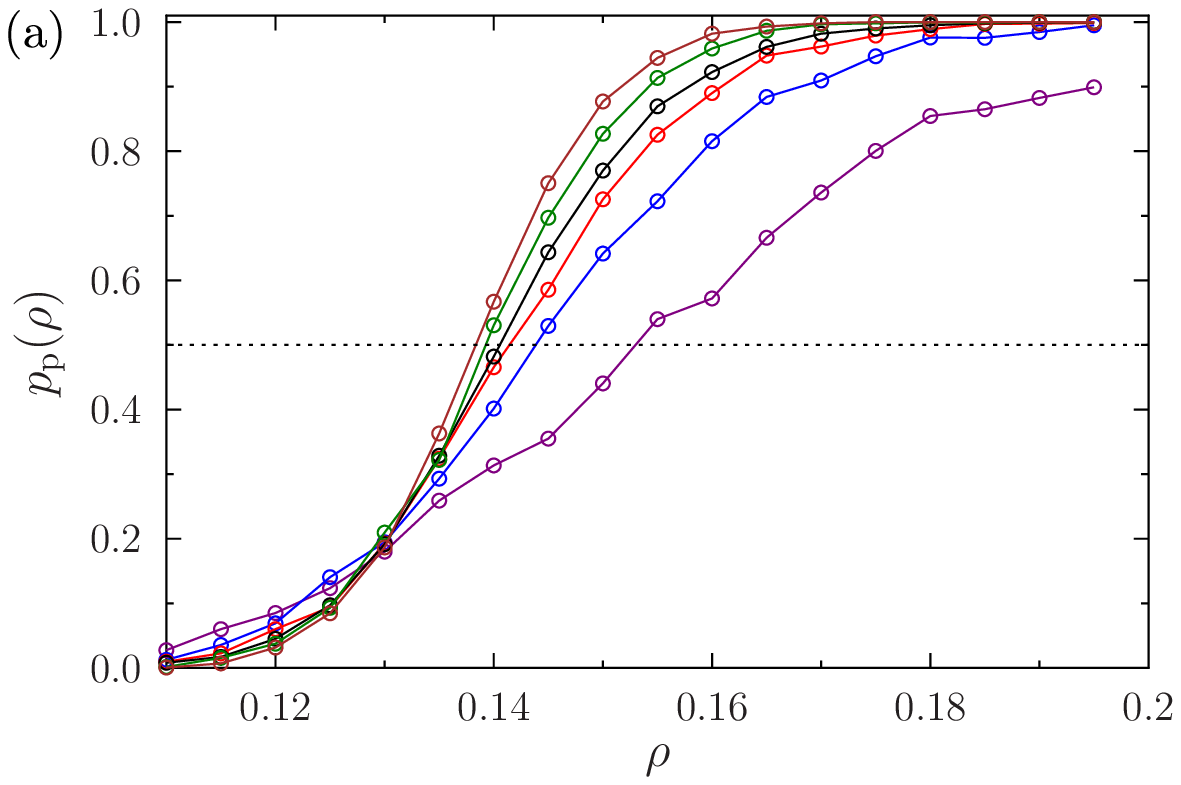}
        \includegraphics[width=0.48\textwidth]{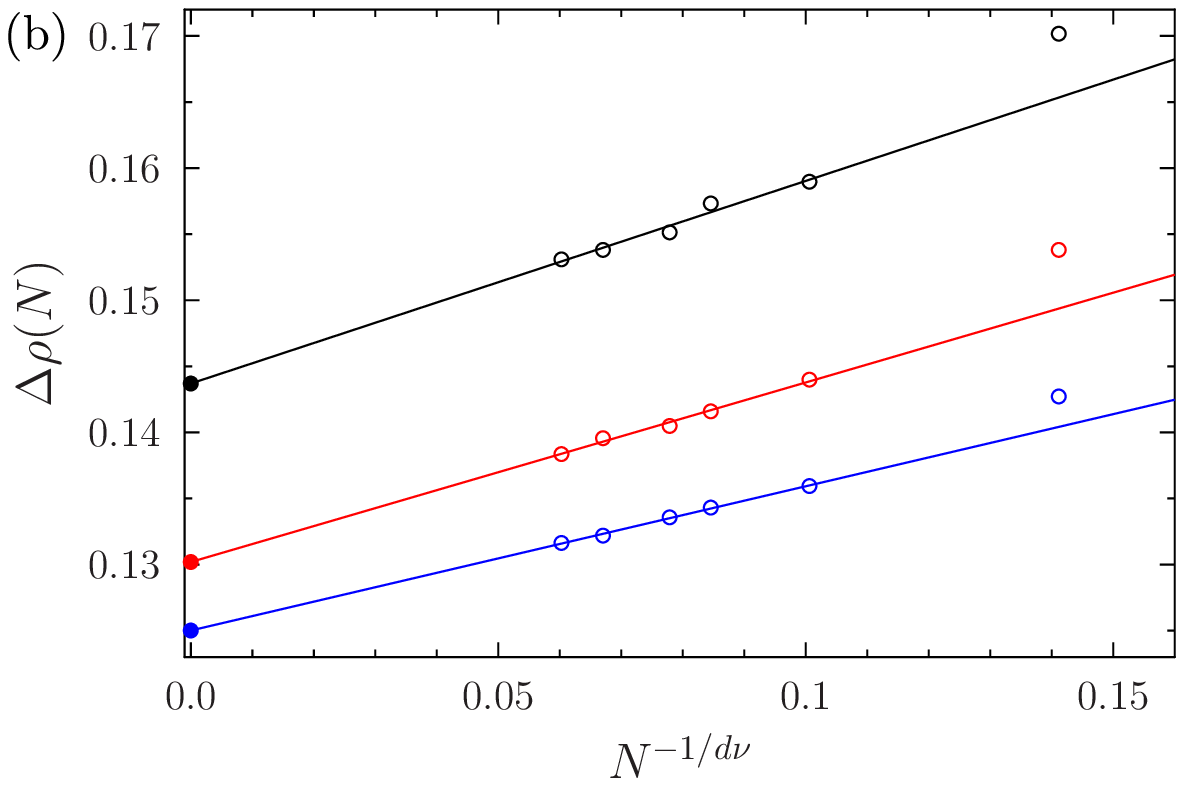}
        \caption{(a) Percolation probability at $T=0.70$ for systems
          of size $N=200$ (purple), $N=500$ (blue), $N=800$ (red),
          $N=1000$ (black), $N=1500$ (green) and $N=2000$ (brown). The
          dashed line indicates the mid-point of the percolation probability, $p_{\frac{1}{2}}$. (b) Extraction of the percolation transition, $\rho_{\mathrm{p}}(T)$ (solid dots), for $T=0.60$ (black), $0.70$ (red) and $0.90$ (purple) from finite-size scaling (empty dots). The solid line is a fit of Eq.~\eqref{eq:perc} to the numerical results.}\label{fig:percolation}
\end{figure}
At low $T$, spherical clusters first turn into wormlike clusters and
then form a disordered percolating network upon increasing $\rho$,
while outside the range of stability of periodic microphases. Changing the cluster fluid morphology to a percolated netowrk dramatically slows down phase-space sampling, which must thence be supplemented by parallel tempering. In order to clearly delineate the hard-to-sample regime, we determine the percolation transition $\rho_{\mathrm{p}}(T)$ by finite-size scaling of the mid-point of the percolation probability
\begin{equation}
\label{eq:perc}
\Delta \rho(N)\equiv|\rho_{\frac12}(N;T)-\rho_{\mathrm{p}}(T)| = N^{-\frac{1}{d\nu}},
\end{equation}
where $d\nu=2.706$ is the universal critical scaling for
three-dimensional standard percolation
(Fig.~\ref{fig:percolation})~\cite{Stauffer1994}. The intercept of the
fit gives the thermodynamic, $N\rightarrow\infty$, percolation
transition. Results are given in the inset of Fig.~4 of the main
text. Note that although the percolation transition can be formally
defined for any $T$, for $T\gtrsim 0.70$, the impact of percolation on
Monte Carlo
sampling is found to be negligible. The inter-particle bonding is then comparable to thermal excitations.

\section{Monte Carlo Simulation Details}
\label{section:simulationDetails}
In order to optimize phase space sampling, we use different sets of Monte Carlo moves for each of the phases.
\begin{itemize}
\item All phases use single-particle local moves following a Metropolis scheme. The maximal displacement is tuned, such that the acceptance ratio is kept between $40\%\text{--}60\%$.
\item All phases also use two types of non-local moves: (i) $10\%$ of the moves are system-wide random displacements, and (ii) $10\%$ of the moves are aggregation volume biased (AVB) moves, which specifically displace a particle to the surface of another~\cite{Chen2000}.  The AVB ``in'' region is set to the attraction radius, i.e., $\sigma<r<\lambda\sigma$, while the  ``out'' region is the rest of the system.
\item Because the first TI step is done at constant pressure, volume changes must be made. For the lamellar morphology, the cartesian directions perpendicular to the lamellae fluctuate together, independently of the third direction. For the cylindrical phase, the ratio between the directions parallel to the cylinder is kept fixed, in order to maintain the cylindrical periodicity. For the other phases, a cubic symmetry of the box is maintained.
\item To accelerate volume sampling of FCC-cluster phase in this same context, we use a cluster volume move algorithm. The move consists of two changes (i) the volume change of equilibrium lattice sites and (ii) the changes between particle, cluster and this lattice sites, see Sec.~\ref{section:clusterVolumeMove}.
\item We use $[N]pT$ simulation and histogram reweighting to determine the coexistence regime near the cluster fluid, FCC-cluster crystal--cylindrical triple point~\cite{Zhang2012}. For each morphology, we iterate the $10^6$ Monte Carlo steps with $5-7$ replicates, in order to ensure sufficient samplings of each configurations.
\item In the cluster fluid regime, $1/N$ of the displacements attempts are virtual move Monte Carlo, which enable cluster displacements~\cite{Whitelam2007}. As the virtual moves have a significant computational overhead, we only use them in the non-percolated cluster fluid regime, where the increase efficiency warrants it.
\item Near and within the percolated fluid regime, parallel tempering is used with attempts at exchanging configurations occuring $1/N$ of the displacements. Temperature intervals are $\Delta T=0.0125$ for $T<0.55$ and $\Delta T=0.0250$ for $T>0.55$. In order to obtain an ergodic sampling, for $\rho<0.25$ the temperature chain goes up to $T=0.70$, while for $\rho>0.45$ the chain goes up to $T=1.00$.
\end{itemize}

Note that in the cluster fluid regime, numerical convergence is checked by determining the free energy by two different approaches for each state point: (i)
Widom insertion at $T=0.70$~\cite{Frenkel2002}, followed by TI over $T$~\cite{Frenkel2002};
and (ii) TI to a targeted $T$, followed by a TI over $\rho$. The pressure is then determined from the
virial, as described in Sec.~\ref{sec:pressure}.  All phase boundaries are determined by common tangent construction from the free energy results.

\subsection{Pressure calculation for SWL Model}
\label{sec:pressure}
In constant volume simulations, the pressure for the SWL model can be determined from its virial
\begin{align}\label{eq:pressure1}
\frac{\beta p}\rho 
&=1 -\frac{2\pi\rho}{3}\int_0^\infty \frac{\ud e^{u(r)}}{\ud r}y(r)r^3\ud r\\
&=1+\frac{2\pi
  \rho}{3}\left\{g(\sigma^+)\sigma^3
    + \left[ g(\lambda\sigma^+)-g(\lambda\sigma^-) \right](\lambda\sigma)^3
+\xi\varepsilon\int_{\lambda\sigma^+}^{\kappa\sigma}g(r)r^2\ud r.
\right\},
\end{align}
recalling that for $r>\sigma$ the radial pair distribution and the cavity functions are equal, i.e., $g(r)=y(r)$.

\subsection{Cluster Volume move algorithm for a multiple-occupancy system }
\label{section:clusterVolumeMove}
In constant pressure simulations of the FCC-cluster phase, a
generalization of the scheme in described in Ref.~\cite{Schultz2011} is
used to enhance sampling. The logarithmic volume move of the lattice sites
\begin{equation}
\log\left(\frac{\br_{\mathrm{lattice,new}}}{\br_{\mathrm{lattice,old}}}\right)=\Delta V^{1/d}
\end{equation}
is then accompanied by changes to the particle positions
\begin{equation}
\frac{\br_{\mathrm{new}}-\br_{\mathrm{lattice,new}}}{\br_{\mathrm{old}}-\br_{\mathrm{lattice,old}}}=\left(\frac{V_{\mathrm{old}}}{V_{\mathrm{new}}}\exp\left[\beta (p(V_{\mathrm{new}}-V_{\mathrm{old}})+\Delta U_{\mathrm{lattice}}\right]\right)^{\frac{1}{d(N-1)}}.
\end{equation}
The acceptance rule is then
\begin{equation}
  \mathrm{acc}(\mathrm{old}\rightarrow\mathrm{new})=\mathrm{min}
  \left\{\exp\left[-\beta (U_{\mathrm{new}}-U_{\mathrm{old}}+\Delta U_{\mathrm{lattice}})\right],1 \right\}.
\end{equation}

\end{widetext}

\end{document}